\documentclass[conference]{IEEEtran}
\IEEEoverridecommandlockouts
\usepackage{cite}
\usepackage{amsmath,amssymb,amsfonts}
\usepackage{algorithmic}
\usepackage{graphicx}
\usepackage{textcomp}
\usepackage{xcolor}
\def\BibTeX{{\rm B\kern-.05em{\sc i\kern-.025em b}\kern-.08em
    T\kern-.1667em\lower.7ex\hbox{E}\kern-.125emX}}
\begin{document}

\title{UKTF: Unified Knowledge Tracing Framework for Subjective and Objective Assessments\\
\thanks{Corresponding author: Chunyan Zeng, Email: cyzeng@hbut.edu.cn.}
}

\author{\IEEEauthorblockN{1\textsuperscript{st} Zhifeng Wang}
	\IEEEauthorblockA{\textit{CCNU Wollongong Joint Institute} \\
		\textit{Central China Normal University}\\
		Wuhan 430079, China \\
		zfwang@ccnu.edu.cn}
	\and
	\IEEEauthorblockN{2\textsuperscript{nd} Jiaqin Wan}
	\IEEEauthorblockA{\textit{Faculty of Artificial Intelligence in Education} \\
		\textit{Central China Normal University}\\
		Wuhan, China \\
		jiaqin\_wan@mails.ccnu.edu.cn}
	\and
	\IEEEauthorblockN{3\textsuperscript{rd} Yang Yang}
	\IEEEauthorblockA{\textit{CCNU Wollongong Joint Institute} \\
		\textit{Central China Normal University}\\
		Wuhan 430079, China \\
		univeryang@ccnu.edu.cn}
	\and
	\IEEEauthorblockN{4\textsuperscript{th} Chunyan Zeng}
	\IEEEauthorblockA{\textit{School of Electrical and Electronic Engineering} \\
		\textit{Hubei University of Technology}\\
		Wuhan 430068, China \\
		cyzeng@hbut.edu.cn}
	\and
	\IEEEauthorblockN{5\textsuperscript{th} Jialiang Shen}
	\IEEEauthorblockA{\textit{Faculty of Artificial Intelligence in Education} \\
		\textit{Central China Normal University}\\
		Wuhan 430079, China \\		
		jialiangshen01@gmail.com}
}



\maketitle

\IEEEpubidadjcol

\begin{abstract}
With the continuous deepening and development of the concept of smart education, learners' comprehensive development and individual needs have received increasing attention. However, traditional educational evaluation systems tend to assess learners' cognitive abilities solely through general test scores, failing to comprehensively consider their actual knowledge states. Knowledge tracing technology can establish knowledge state models based on learners' historical answer data, thereby enabling personalized assessment of learners. Nevertheless, current classical knowledge tracing models are primarily suited for objective test questions, while subjective test questions still confront challenges such as complex data representation, imperfect modeling, and the intricate and dynamic nature of knowledge states. Drawing on the application of knowledge tracing technology in education, this study aims to fully utilize examination data and proposes a unified knowledge tracing model that integrates both objective and subjective test questions. Recognizing the differences in question structure, assessment methods, and data characteristics between objective and subjective test questions, the model employs the same backbone network for training both types of questions. Simultaneously, it achieves knowledge tracing for subjective test questions by universally modifying the training approach of the baseline model, adding branch networks, and optimizing the method of question encoding. This study conducted multiple experiments on real datasets, and the results consistently demonstrate that the model effectively addresses knowledge tracing issues in both objective and subjective test question scenarios.
\end{abstract}


\section{Introduction}
With the thriving development of online adaptive learning platforms \cite{Li2023e,Wang2023h}, educational information technology is gradually infiltrating every aspect of education \cite{Liao2024}. In this context, knowledge tracing technology \cite{Wang2024i,Li2023g,Wang2023c}, as the cornerstone of online intelligent education, bears the crucial task of providing precise support for personalized learning, thereby attracting continuous attention from researchers in the field of educational technology \cite{Lyu2022}.

In 1972, Atkinson first introduced the knowledge tracing model\cite{r1}, a seminal model simulating learners' mastery of knowledge. Subsequently, in 1995, Corbett and Anderson incorporated the knowledge tracing model into the realm of intelligent education and successfully applied it to intelligent educational systems\cite{r2}. Early knowledge tracing models primarily relied on machine learning methods, which, despite their simplicity and interpretability, often yielded unsatisfactory results. Recently, with the rise of deep learning \cite{Wang2023k}, an increasing number of deep learning approaches have been applied to the field of knowledge tracing. Leveraging its powerful feature representation and learning capabilities, deep learning has provided more precise and effective modeling tools for knowledge tracing.

This study employs three classic models as baselines: the Deep Knowledge Tracing model (DKT), the Dynamic Key-Value Memory Networks model (DKVMN), and the Graph-based Knowledge Tracing model (GKT). Drawing on learners' performance on both objective and subjective test questions, we have realized a unified knowledge tracing model for both types of questions. The primary contributions of this research are as follows:

1) Given the binary nature of answers to objective test questions, the model adopts a classification approach for training, utilizing the backbone network to trace knowledge in objective test questions. As subjective test scores follow a multi-value discrete distribution, to accommodate the assessment and prediction requirements of subjective test questions, we convert the binary classification problem into a regression problem. 

2) While maintaining a unified backbone network, we achieve knowledge tracing for subjective test questions by uniformly altering the model's training approach, adding branch networks, and optimizing question encodings. 

3) This study conducted extensive experiments using realworld datasets, demonstrating that the model effectively addresses knowledge tracing issues in both objective and subjective test question scenarios.

The rest of the paper is organized as follows. In Section \ref{RW}, we review the related work. In Section \ref{MED}, we introduce the main research methods. In Section \ref{EXP}, we describe the details of the experiment and the results. Finally, we have a summary of this work in Section \ref{CON}.

\section{Related Work} \label{RW}
Current knowledge tracing models are primarily divided into two categories: traditional knowledge tracing models based on machine learning and knowledge tracing models based on deep learning.

\subsection{Traditional Knowledge Tracing Models Based on Machine Learning}
Although traditional knowledge tracing models based on machine learning offer good interpretability, they rely on theoretical assumptions and require manual construction of input features, which can often be one-sided and limited. Consequently, the predictive performance of these models is generally mediocre. Bayesian Knowledge Tracing (BKT) is a representative model from this period. Its core principle is based on the Hidden Markov Model (HMM) for time series analysis \cite{Li2023g}. However, early BKT model simply categorized students' knowledge states into two categories: mastered and not mastered. To address this issue, Zhang et al. \cite{r3} extended the original two states to three, including an intermediate possible mastery state. Wang et al. \cite{r4}refined the representation of students' learning status by replacing binary nodes with continuous values between 0 and 1. Given the potential dependencies between knowledge concepts, K¨aser proposed Dynamic Bayesian Knowledge Tracing (DBKT) \cite{r5}. In the DBKT model, a learner's mastery level of a particular knowledge concept is also constrained by their mastery of other related concepts.

\subsection{Knowledge Tracing Models Based on Deep Learning}
Deep neural networks have been a huge success in speech \cite{barolli_deep_2021,zhu_liveness_2013,wang_playback_2011} and image processing \cite{wang_sae_2021,tian_occlusion_2018}, and knowledge tracking as well. Based on the different neural networks used, knowledge tracing models based on deep learning can be broadly classified into four categories: those based on Recurrent Neural Networks (RNN) \cite{r6}, those based on Dynamic Key-Value Memory Networks \cite{r7}, those incorporating Attention Mechanisms\cite{r8}, and those utilizing Graph Neural Networks (GNN)\cite{r9}. Among these, the DKT, DKVMN, and GKT models are the focus of this study.

The Deep Knowledge Tracing model (DKT), pioneered by Chris Piech et al. in 2015 \cite{r10}, introduced a novel modeling approach for the field of knowledge tracing using Recurrent Neural Networks (RNN)\cite{r6}. Addressing the issues of input irreconstructibility and persistent fluctuations in learner knowledge levels in the DKT model, the DKT+ model\cite{r11} optimized the loss function by introducing regularization terms to ensure the stability of student knowledge levels. In 2017, Zhang et al. combined the advantages of Memory Augmented Networks (MANN) to propose the Dynamic Key-Value Memory Networks (DKVMN) \cite{r7}, which, for the first time, employed a dynamic key-value memory network approach to enable modeling of multiple knowledge concept levels. While DKVMN improved interpretability compared to DKT, it struggled to effectively capture sequential dependencies in learners' answer sequences. To address this, Abdelrahman et al.\cite{r12} proposed the SKVMN model, which introduced an improved Hop-LSTM neural network to tackle the issue of DKVMN's inability to capture long-term dependencies in learner interactions. In 2019, Nakagawa et al. \cite{r13} first applied Graph Neural Networks to knowledge tracing, introducing the GKT model, which utilizes a graph structure to represent the relationships between students' answer histories and knowledge concepts. The GKT model reformulates knowledge tracing as a temporal node-level classification problem in GNN, decomposing course knowledge into several knowledge concepts and utilizing GNN for node state updates and information aggregation to predict students' future mastery of various knowledge points. Subsequently, Yang et al. further introduced the GIKT model \cite{r14}, which fully integrates the correlation between problems and skills through embedding propagation.

\subsubsection{Deep Knowledge Tracing Model}

\begin{itemize}
\item \textit{Answer Data Modeling}: The learner's answer data is organized into a time series, where each time point corresponds to an answer event. Each answer event comprises question information and the answer result. The learner's answer data at time $t$, denoted as $x_t$, is set as a one-hot encoding of the learner's interaction tuple $\{e_t, r_t\}$, where $x_t \in \{0,1\}^{2N}$.

\item \textit{Knowledge State Modeling}: The Long Short-Term Memory model (LSTM) is employed to model the temporal dependencies in learners' answer sequences. A knowledge state variable $h_t$ is introduced to represent the learners' mastery level of the knowledge concepts.

\item \textit{Parameter Learning and Optimization}: Through supervised learning, the parameters of the model are learned. Using known answer data, the parameters are adjusted through the backpropagation algorithm and a gradient descent optimizer to minimize the difference between the predicted answer outcomes and the actual answer outcomes. The loss computation for the entire answer sequence of a single learner is as follows:
\begin{equation}
    L=\sum_t\ell(y^T\delta(e_{t+1}),r_{t+1})
\end{equation}
\item \textit{Answer Prediction}: The knowledge state variable $h_t$ is passed through an output layer to map it to the probability of the student answering each question correctly, thereby enabling the prediction of answers.
\end{itemize}

\subsubsection{The Dynamic Key-Value Memory Network Model}
\begin{itemize}
\item \textit{Knowledge Encoding}: The learner's answer data is encoded into a form suitable for model processing, and firstly, one-hot encoding is applied to obtain the neural network's input data $e_t$ and the combined value matrix components' data $x_t = (e_t, r_t)$, where $q_t$ represents the question label at time $t$, and $r_t$ represents the learner's response at $t$. Subsequently, an embedding layer is utilized to encode the data into continuous embedding vectors $m_t$ for $e_t$ and $s_t$ for the combination $(e_t, r_t)$.

\item \textit{Knowledge State Modeling}: The DKVMN network is employed to model the changes in students' knowledge states. Two memory matrices are introduced: a static key matrix and a dynamic value matrix. The key matrix represents knowledge concepts, while the value matrix represents the learners' mastery levels of these concepts.

\item \textit{Model Training and Optimization}: During the training process, the standard cross-entropy loss between the predicted answer probability $p_t$ and the true label $l_t$ is minimized using the gradient descent algorithm to jointly learn the embedding matrices and other parameters. The formula for the loss function is as follows:  
\begin{equation}
    L=-\sum_t (l_t\mathrm{log}p_t+(1-l_t)\log(1-p_t))
\end{equation}
\item \textit{Answer Prediction}: By incorporating the vector $f_t$ that captures the learner's comprehensive knowledge level and question characteristics, the probability $p_t$ of the learner correctly answering the question is obtained through a fully connected layer with a sigmoid activation function, thereby enabling prediction of the learner's response.  Specifically, it is defined as: 
\begin{equation}
    p_t=\sigma\left(W_f\cdot f_t+b_f\right)
\end{equation}
\end{itemize}

\subsubsection{The Graph-based Knowledge Tracing Model}

\begin{itemize}
\item \textit{Knowledge Graph Construction}: The GKT model structures the course exercises into a graph $G=(V,E,A)$ potentially, decomposing the requirements for mastering course exercises into $N$ knowledge concepts, referred to as the node set $V=\{v_1,\cdots,v_N\}$. These knowledge concepts share dependencies, denoted as edges $E \subseteq V \times V$, and the degree of dependency is represented by the adjacency matrix $A \in \mathbb{R}^{N \times N}$.

\item \textit{Knowledge State Modeling}: The learner, at time step $t$, possesses an independent knowledge state $h^t = \{h_i^t \mid i \in V\}$ for each knowledge concept, where this knowledge state evolves over time. When the learner attempts an exercise $v_i$ related to knowledge concept $i$, not only does the learner's knowledge state for concept $i$ update, but also the knowledge states $h_j^t$ of adjacent concepts $j \in N_i$ are updated, where $N_i$ denotes the set of nodes adjacent to node $v_i$. The updating process in the GKT model is accomplished through two steps: Aggregation and Update, performed by a Graph Convolutional Network (GCN).

\item \textit{Model Training and Optimization}: During the training process, the model is trained by minimizing the Negative Log-Likelihood (NLL) loss in order to achieve the best model performance.
\item \textit{Answer Prediction}: Based on the updated knowledge states, the prediction probability $p_t$ of the learner correctly answering each knowledge concept is obtained through a fully connected layer with a sigmoid activation function, enabling the prediction of the learner's responses to questions.
\end{itemize}

\section{Proposed Methods} \label{MED}
\begin{figure}[htbp]
\centerline{\includegraphics[width=0.5\textwidth]{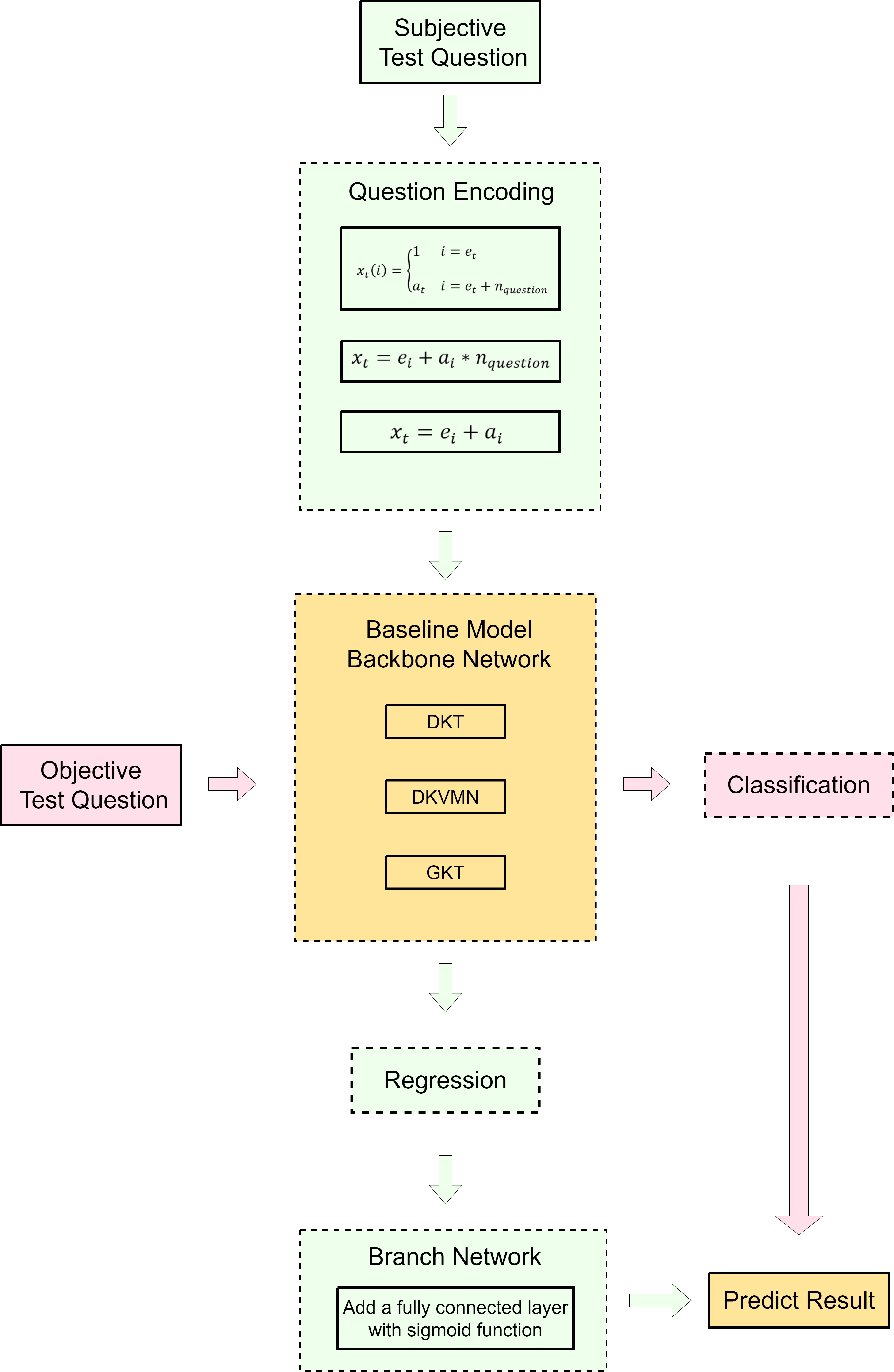}}
\caption{A Unified Knowledge Tracing Model for Both Objective and Subjective Test Questions.}
\label{fig2}
\end{figure}

The specific framework of the unified knowledge tracing model for both objective and subjective test questions is shown in Fig. \ref{fig2}. To construct the proposed model, this study selects three classic models in the field of knowledge tracing as backbones: the Deep Knowledge Tracing Model (DKT), the Dynamic Key-Value Memory Network Model (DKVMN), and the Graph-based Knowledge Tracing Model (GKT). The models are conducted using learners' answer data on both objective and subjective test questions.

\subsection{Knowledge Tracing Models in Objective Test Question Scenarios}\label{AA}
\subsubsection{Deep Knowledge Tracing Model}
DKT model predicts learners' future performance in learning by analyzing their historical answer data, employing Long Short-Term Memory (LSTM) to capture the temporal changes in learners' knowledge states \cite{Li2023b}. The specific definitions are as follows:
\begin{equation}
    h_t=\tanh(\mathrm{W}_{hx}\mathrm{x}_t+\mathrm{W}_{hh}\mathrm{h}_{t-1}+\mathrm{b}_h)
\end{equation}
\begin{equation}
    y_t=\sigma\big(\mathrm{W}_{yh}\mathrm{h}_t+\mathrm{b}_y\big)
\end{equation}
Where $h_{t-1}$ represents the hidden state at the previous time step, $h_t$ denotes the current knowledge state of the learner, and $y_t$ represents the probability of the learner correctly answering at time $t$.
\subsubsection{Dynamic Key-Value Memory Networks Model}
DKVMN model takes learners' answer data as input and utilizes a dynamic key-value memory matrix to store learners' knowledge states. The model employs two memory matrices: a static key matrix and a dynamic value matrix, which are used to store fixed knowledge point information and learners' dynamically changing knowledge states, respectively. The DKVMN model utilizes an attention mechanism to control read and write operations, thereby enabling dynamic updates to the knowledge state. It is primarily composed of three components:
\begin{itemize}
\item \textit{Acquire Relevant Weights}: The query is performed by leveraging the embedding vector $m_t$ to interrogate the key storage matrix $M^k$. This involves computing the inner product between $m_t$ and each key slot $M^k_{(i)}$ within the matrix. Subsequently, a softmax activation function is applied to transform these inner products into a probability distribution, representing the weights $w_t \in \mathbb{R}^N$ that quantify the involvement of each knowledge point in the question at time $t$. The computation formula is as follows:  
\begin{equation}
    w_t(i)=\mathrm{Softmax}(m_t^TM^k(i))
\end{equation}
\item \textit{Reading Process}: The reading vector $r_t$ is computed based on the value matrix $M_t^v$ and the knowledge point relevance weights $w_t$. The vector $r_t$ represents the learner's mastery level of the given exercise. The computation formula is as follows:  
\begin{equation}
    r_t=\sum_{i=1}^Nw_t(i)M_t^v(i)
\end{equation}

Given that each exercise possesses its inherent difficulty, the reading vector $r_t$ and the embedding $m_t$ of the input exercise are concatenated vertically. This concatenated vector is then passed through a fully connected layer with a tanh activation function, yielding an aggregated feature vector $f_t$, which encapsulates both the learner's mastery level and the prior difficulty of the exercise. Subsequently, $f_t$ is fed into another fully connected layer with a sigmoid activation function to predict the probability $p_t$ that the learner will correctly answer the question. The specific implementation process is outlined as follows:  

\begin{equation}
    f_t=\tanh(W_1^T[r_t,m_t]+b_1)
\end{equation}
\begin{equation}
    p_t=\mathrm{sigmoid}(W_2^Tf_t+b_2)
\end{equation}

\item \textit{Writing Process}: After the learner answers the question, the value memory matrix $M_t^v$ is updated based on the learner's response. When incorporating the learner's knowledge growth $s_t$ into the value matrix components, the existing memory is first erased before adding the new information.  
  
Given a write weight $w_t$ (which is the same as the knowledge point relevance weight $w_t$ utilized in the reading process), the computation of the erase vector $e_t$ is performed as follows:  
\begin{equation}
    e_t=\mathrm{sigmoid}(E^Ts_t+b_e)
\end{equation}
Where the transformation matrix $E \in \mathbb{R}^{d_v \times d_v}$ has a shape of $d_v \times d_v$, and $e_t \in \mathbb{R}^{d_v}$ is a column vector with $d_v$ elements, each element taking a value within the interval $(0, 1)$. The memory vector of the value component $M_{t-1}^v(i)$ from the previous time step is modified as follows:  
\begin{equation}
    \tilde{M}_t^v(i)=M_{t-1}^v(i)[\mathbf{1}-w_t(i)e_t]
\end{equation}
Where $\mathbf{1}$ represents a row vector of all ones, and thus, only when both the weight at the corresponding matrix position and the erase element are 1, will the element at that matrix position be reset to 0. 
After erasing the memory, an add vector $a_t$ is employed to update the value matrix $M_t^v$, and the update of the value matrix at each time step is given by:  
\begin{equation}
    M_t^v(i)=\tilde{M}_{t-1}^v(i)+w_t(i)a_t
\end{equation}
Given the transformed memory information and the current problem features, a feature vector $f_t$ is constructed. Subsequently, this feature vector is fed into a fully connected layer with a sigmoid activation function to predict the probability $p_t$ that the learner will correctly answer the question. The specific implementation process is detailed as follows:  
\begin{equation}
    p_t=\mathrm{sigmoid}(W_2^Tf_t+b_2)
\end{equation}
\end{itemize}

\subsubsection{The Graph-based Knowledge Tracing Model}
The GKT model \cite{r13} represents learners' historical answer data as a graph structure, where each node signifies a knowledge concept, and the edges between nodes indicate the relationships between these knowledge concepts. By observing learners' answer patterns, the model updates the states of the nodes within the graph, thereby dynamically tracking learners' mastery levels of various knowledge concepts. This model primarily comprises two components:
\begin{itemize}
\item \textit{Information Aggregation}: We define an answer embedding matrix $E_x \in \mathbb{R}^{2N \times e}$ and a skill embedding matrix $E_c \in \mathbb{R}^{N \times e}$, where $E_c(k)$ represents the $k$-th row vector of matrix $E_c$, and $e$ is the embedding dimension. At time step $t$, for the answered knowledge concept $i$ and its neighboring knowledge concepts $j$, the hidden states and embeddings are aggregated according to the following specific formula:  
\begin{equation}
    h_k^{\prime t} =   
    \begin{cases}  
    [h_k^t, x^t E_x] &  (k = i) \\  
    [h_k^t, E_c(k)] &  (k \neq i) 
    \end{cases}  
\end{equation}
\item \textit{Feature Update}: Update the hidden state of each knowledge point through aggregated features and the structure of the knowledge graph. The specific steps are as follows:
\begin{equation}
    m_k^{t+1}=
    \begin{cases}
    f_\text{self}(h_k^{\prime t})&(k=i)\\
    f_\text{neighbor}(h_i^{\prime t},h_k^{\prime t})&(k\neq i)
    \end{cases}
\end{equation}
\begin{equation}
    \widetilde{m}_k^{t+1}=G_{ea}(m_k^{t+1})
\end{equation}
\begin{equation}
    h_k^{t+1}=G_{gru}(\tilde{m}_k^{t+1},h_k^t)
\end{equation}

In this context, $f_{\text{self}}$ and $f_{\text{neighbor}}$ represent the self-function and neighbor-function respectively, where $f_{\text{self}}$ is implemented as a Multilayer Perceptron (MLP). $G_{\text{ea}}$ denotes an Erase-Add gate, which performs feature erasing and enhancement on the aggregated feature vector $m_k^{(t+1)}$ to better simulate the learner's knowledge state characteristics. Additionally, $G_{\text{gru}}$ is a Gated Recurrent Unit (GRU) that is utilized to perform feature extraction, memory retention, and updating on $\tilde{m}_k^{(t+1)}$.  

$f_{neighbor}$ is an arbitrary function defined based on the structure of the knowledge graph to propagate information to adjacent nodes. Nakagawa et al. proposed two optional implementations: a statistical-based approach and a learning-based approach. In this study, a statistical-based dense graph approach is adopted, with the formula outlined as follows:  
\begin{equation}
    f_{\mathrm{outgo}}(\mathrm{h'}_i^t,\mathrm{h'}_j^t)=\mathbf{A}_{i,j}f_{\mathrm{outgo}}([\mathrm{h'}_i^t,\mathrm{h'}_j^t])
\end{equation}
\begin{equation}
    f_{\mathrm{income}}(\mathrm{h'}_i^t,\mathrm{h'}_j^t)=\mathbf{A}_{j,i}f_{\mathrm{income}}([\mathrm{h'}_i^t,\mathrm{h'}_j^t])
\end{equation}
\begin{equation}
    f_{\mathrm{neighbor}}=f_{\mathrm{outgo}}(\mathrm{h'}_i^t,\mathrm{h'}_j^t)+f_{\mathrm{income}}(\mathrm{h'}_i^t,\mathrm{h'}_j^t)
\end{equation}
The dense graph is computed as follows:
\begin{equation}
    \mathbf{A}_{i,j}=
    \begin{cases}
    \frac{n_{i,j}}{\Sigma_{k}n_{i,k}}&\quad\mathrm{i\neq j}\\
    \\0&\quad\mathrm{i=j}
    \end{cases}
\end{equation}

Based on the updated knowledge state, the prediction probability $p_t$ of the learner correctly answering each knowledge point at the next time step is obtained through a fully connected layer with a sigmoid activation function, thereby enabling prediction of the learner's responses. The specific formula is as follows:  
\begin{equation}
    p_k^t=\sigma(W_\text{out}h_k^{t+1}+b_k)
\end{equation}
\end{itemize}

\subsection{Knowledge Tracing Models in Subjective Test Question Scenarios}
\subsubsection{Deep Knowledge Tracing Model}
Since the scores of subjective test questions are multi-valued discretely distributed rather than simply right or wrong. Therefore, some of the structure of the DKT model needs to be adjusted to fit the new prediction task. The specific adjustments are as follows:
\begin{itemize}
\item \textit{Question Encoding}: For the response tuple $x_t = (e_t, a_t)$, after one-hot encoding, $x_t \in \{0,1\}^{2E}$, with a length twice the total number of exercise items. For subjective test questions, this experiment directly encodes the question ID and corresponding score into the feature vector. Specifically, if an exercise item is attempted, the corresponding position in the first half of $x_t$ is set to 1, otherwise 0; if an exercise item is attempted, the corresponding position in the second half of $x_t$ represents the specific score, otherwise 0. The detailed encoding formula is as follows:  
\begin{equation}
    x_t(i)=\begin{cases}1&i=e_t\\
    a_t&i=e_t+n_{question}
    \end{cases}
\end{equation}

\item \textit{Loss Function}: For the subjective test questions, the knowledge-tracking task can be defined simply as a regression problem, so the loss function was adjusted from BCELoss to MSELoss.
\item \textit{Branch Network}: The final output layer of the original DKT model usually uses a softmax activation function to predict how well the learner will answer the questions, right or wrong. For subjective test questions, a fully connected layer with a sigmoid function will be added to output the learners' scores on the subjective test questions.
\end{itemize}

\subsubsection{Dynamic Key-Value Memory Networks Model}
In response to the scoring characteristics of the subjective questions, some adjustments to the structures of the DKVMN model are as follows:
\begin{itemize}
\item \textit{Question Encoding}: For subjective test items, the problem IDs are encoded using one-hot encoding, and the response tuple $x_t = (e_t, a_t)$ is correlated with the number of knowledge points, where the specific encoding formula is given as follows:
\begin{equation}
    x_t=e_i+a_i*n_{question}
\end{equation}
\item \textit{Loss Function}: The original DKVMN model picks \emph{binary cross-entropy with logits} for model optimization. For the subjective test questions, the loss function was adjusted to MSELoss.

\item \textit{Branch Network}: The final output of the original DKVMN model is the probability of answering the test question correctly, and for the subjective test questions, a fully connected layer with a sigmoid function is added to change the output to a normalized score for the test question.
\end{itemize}

\subsubsection{The Graph-based Knowledge Tracing Model}
The experiment adapted the GKT model by modifying the classification task to a regression task and applying the GKT model to the prediction of scores on subjective test questions.
\begin{itemize}
\item \textit{Question Encoding}: The encoding process fuses the knowledge point ID $e_t$  with the answer situation $a_t$ into a single vector, with a specific encoding formula as follows:
\begin{equation}
    x_t=e_i+a_i
\end{equation}
\item \textit{Loss Function}: The negative likelihood logit (NLL) was adjusted to MSELoss, and the GKT model was adjusted to be able to predict subjective test scores.
\end{itemize}

\section{Experimental Results and Analysis} \label{EXP}
\subsection{Dataset}
In the objective test question scenario, each model was applied on three real datasets. The datasets used for the experiments are the publicly available dataset \textit{ASSISTment-2009-2010-skill} dataset, and the monthly test answer dataset of the sophomore class of Changshui Middle School, which contains objective test question data of eight academic disciplines, abbreviated as \textit{Assist09} and \textit{Object}.

\textit{Assist09} is an ensemble of learners' practice records for the 2009-2010 school year provided by the ASSISTments online tutoring platform, which is a widely used dataset in knowledge tracking. This dataset has two data formats:\textit{ “Skill-builder”} and\textit{ “Non-Skill-builder”}. In this experiment, the data set in \textit{“Skill-builder”} format is selected for the experiment. \textit{Object} comes from the monthly examination records of the sophomore class of Changshui Middle School, which contains the objective test data of eight academic disciplines.

\begin{table}[htbp]
\caption{Statistical information about Datasets}
\begin{center}
\begin{tabular}{c|c|c|c|c}
\hline
Dataset           & Student  & Skill & Question   & Exercise \\ \hline
Assist09 & 4217 & 124 & 26688 & 401756\\ \hline  
Object-English & 4773 & 23 & 65 & 310245 \\ \hline  
Object-Math & 5224 & 16 & 16 & 83584 \\ \hline  

\end{tabular}
\label{tab1}
\end{center}
\end{table}

\begin{table}[htbp]
\caption{Statistical information about Datasets}
\begin{center}
\begin{tabular}{c|c|c|c|c}
\hline
Dataset           & Student  & Skill & Question   & Exercise \\ \hline
Subject-Chinese & 5236 & 21 & 24 & 125664 \\ \hline
Subject-Math    & 5224 & 6  & 6  & 31344  \\ \hline
Subject-Biology & 2901 & 5  & 5  & 14505  \\ \hline
\end{tabular}
\label{tab2}
\end{center}
\end{table}

In the objective test scenario, taking the DKT model experiment as an example, we introduce the three datasets used in the experiment, and the raw statistical information is shown in Table \ref{tab1}. 

In the subjective test scenario, the experiment uses Subject, a monthly test answer dataset of the sophomore class of Changshui Middle School, which contains subjective test data of eight subjects.
Taking the DKT model experiment as an example, the three datasets used for the experiment are introduced, and the raw statistical information is shown in Table \ref{tab2}.

\begin{table}[htbp]
\caption{Description of  \textit{Assit09}}
\begin{center}
\begin{tabular}{c|c}
\hline
Name      & Meaning  \\ \hline
order\_id & Non-time-series aligned exercise record ID   \\ \hline
user\_id  & Student ID    \\ \hline
skill\_id & Concept ID    \\ \hline
correct   & Students' answering questions correctly or incorrectly \\ \hline
\end{tabular}
\label{tab3}
\end{center}
\end{table}

\begin{table}[htbp]
\caption{Description of  \textit{Object}}
\begin{center}
\begin{tabular}{c|c}
\hline
Name      & Meaning  \\ \hline
exer\_id & Non-time-series aligned exercise record ID   \\ \hline
user\_id  & Student ID    \\ \hline
knowledge\_code & Concept ID    \\ \hline
score   & Students' score \\ \hline
\end{tabular}
\label{tab4}
\end{center}
\end{table}

Four fields related to the model are selected from the raw data of the three datasets, taking \textit{ASSISTment-2009-2010-skill} as an example, the specific meanings are shown in Table \ref{tab3}. Taking \textit{Object} as an example, the specific meaning is shown in Table \ref{tab4}.

\subsection{Data Pre-processing}
In the objective test question scenario, taking \textit{ASSISTment-2009-2010-skill} as an example, the following preprocessing is done on the extracted data:
\begin{itemize}
\item For all records containing missing values (NaN) in the skill\_id column, the deletion operation is performed.
\item For learner answer records containing the same interaction data, delete operation is performed and only one piece of duplicate data is retained.
\item Convert the question ID to a continuous variable starting from 0.
\item Convert data samples into sequential data categorized by ID.
\item Make one-hot encoding of question IDs and answers.
\item Uniform sequence length
\item For subjective test questions, the process of normalizing test scores also needs to be added. Normalized scores are on similar scales and are easier to compare and interpret.
\end{itemize}

\subsection{Experiment Result}
\begin{itemize}
\item \textit{Model Performance in Objective Test Question Scenarios}
\end{itemize}

The DKT model uses LSTM structure, Adam trainer to predict the learners' performance in answering the objective test questions for the next moment, and calculates BCELoss, RMSE, MAE, and AUC in each Epoch by comparing the real scores with the predicted scores. The DKT model is experimented on three datasets, and the model performs well in all of them.

\begin{table}[htbp]
\caption{DKT Performance on objective test questions}
\begin{center}
\begin{tabular}{c|c|c|c|c}
\hline
Dataset        & BCELoss & RMSE  & MAE   & AUC   \\ \hline
Assit09        & 0.503   & 0.407 & 0.329 & 0.798 \\ \hline
Object-English & 0.574   & 0.442 & 0.392 & 0.763 \\ \hline
Object-Math    & 0.491   & 0.404 & 0.329 & 0.816  \\ \hline
\end{tabular}
\label{tab5}
\end{center}
\end{table}
As shown in Table \ref{tab5}, the AUC of all three datasets reached over 0.75, which can prove that the DKT model is effective on the objective test knowledge tracking task.

Among them, the DKT model performs best on the dataset \textit{Object-Math}. Analyzed in combination with the size and complexity of the dataset, the data structure of \textit{Object-Math} is simpler and the DKT model fits better. When the data structure is complex but the amount of data is not sufficient, such as the dataset \textit{Object-English}, the prediction performance is degraded to some extent.


In order to evaluate the prediction effect of DKVMN model on objective test questions, this paper conducted experiments on three real datasets, and at the same time, four commonly used metrics, namely, Loss, RMSE, ACC, and AUC, were chosen as the metrics for evaluating the model.

\begin{table}[htbp]
\caption{DKVMN Performance on objective test questions}
\begin{center}
\begin{tabular}{c|c|c|c|c}
\hline
Dataset        & Loss  & RMSE  & ACC   & AUC   \\ \hline
Assit09        & 0.501 & 0.494 & 0.755 & 0.800 \\ \hline
Object-English & 0.468 & 0.486 & 0.763 & 0.834 \\ \hline
Object-Math    & 0.430 & 0.458 & 0.790 & 0.862  \\ \hline
\end{tabular}
\label{tab6}
\end{center}
\end{table}

As shown in Table \ref{tab6}, the AUC scores on the dataset were all above 0.75 and the ACC scores were all above 0.80, proving that the DKVMN model has good generalization ability on the objective test knowledge tracking task. 

Among them, the DKVMN model fits best on the dataset \textit{Object-Math}, with a Loss of 0.430, an RMSE score of 0.458, an ACC score of 0.790, and an AUC score of 0.862.


The GKT model chooses the negative log-likelihood (NLL) as a measure of Loss, while three commonly used metrics, RMSE, ACC, and AUC, are chosen to evaluate the model performance. 
\begin{table}[htbp]
\caption{GKT Performance on objective test questions}
\begin{center}
\begin{tabular}{c|c|c|c|c}
\hline
Dataset        & Loss  & RMSE  & ACC   & AUC   \\ \hline
Assit09        & 0.562 & 0.435 & 0.718 & 0.723 \\ \hline
Object-Math    & 0.466 & 0.393 & 0.768 & 0.845 \\ \hline
Object-Biology & 0.531 & 0.423 & 0.723 & 0.770  \\ \hline
\end{tabular}
\label{tab7}
\end{center}
\end{table}

As shown in Table \ref{tab7}, the AUC scores and ACC scores on the dataset are all above 0.7, and the results indicate that the GKT model is able to fit a wide range of different learner data for the knowledge tracking task. 

Among them, the GKT model fits best on the dataset \textit{Object-Math}, with Loss up to 0.466, RMSE score up to 0.393, ACC score up to 0.768, and AUC score up to 0.845.

\begin{itemize}
\item \textit{Model Performance in Subjective Test Question Scenarios}
\end{itemize}

Experiments were carried out with the DKT model on the three subjective test question datasets to calculate MSELoss, RMSE, MAE, and ACC in each Epoch, and the optimal model performance on the three datasets was as follows:
\begin{table}[htbp]
\caption{DKT Performance on subjective test questions}
\begin{center}
\begin{tabular}{c|c|c|c|c}
\hline
Dataset         & MSELoss & RMSE  & MAE   & ACC   \\ \hline
Subject-Chinese & 0.105   & 0.325 & 0.259 & 0.844 \\ \hline
Subject-Math    & 0.086   & 0.294 & 0.226 & 0.896 \\ \hline
Subject-Biology & 0.040   & 0.201 & 0.165 & 0.905  \\ \hline
\end{tabular}
\label{tab8}
\end{center}
\end{table}

As shown in Table \ref{tab8}, the ACC on all three datasets reached above 0.8, which can prove that the improved DKT model is effective on the subjective test knowledge tracking task. Among them, the model performs best on the dataset \textit{Subject-Biology}, with ACC score up to 0.905.


The final output of the improved DKVMN model is the normalized scores of the subjective questions, and the four commonly used metrics of MSELoss, RMSE, MAE, and ACC are chosen as the metrics to evaluate the model in this experiment. The optimal model performance on the three datasets is as shown in Table \ref{tab9}:

\begin{table}[htbp]
\caption{DKVMN Performance on subjective test questions}
\begin{center}
\begin{tabular}{c|c|c|c|c}
\hline
Dataset         & MSELoss & RMSE  & MAE   & ACC   \\ \hline
Subject-Math    & 0.057   & 0.316 & 0.263 & 0.686 \\ \hline
Subject-Biology & 0.039   & 0.309 & 0.260 & 0.718 \\ \hline
Subject-Physics & 0.093   & 0.362 & 0.295 & 0.704  \\ \hline
\end{tabular}
\label{tab9}
\end{center}
\end{table}

From the data in the table, it is clear that the improved DKVMN model is effective on the subjective test knowledge tracking task. 

Among them, the model performs best on the dataset \textit{Subject-Biology} and fits less well on the dataset \textit{Subject-Math}.


The experimental datasets of the GKT model are \textit{Subject-Chinese}, \textit{Subject-Math}, and \textit{Subject-Physics}.The performance of the adjusted GKT model on the three datasets is shown in Table \ref{tab10}:

\begin{table}[htbp]
\caption{GKT Performance on subjective test questions}
\begin{center}
\begin{tabular}{c|c|c|c|c}
\hline
Dataset         & MSELoss & RMSE  & MAE   & ACC   \\ \hline
Subject-Chinese & 0.106   & 0.326 & 0.195 & 0.799 \\ \hline
Subject-Math    & 0.041   & 0.202 & 0.041 & 0.960 \\ \hline
Subject-Physics & 0.133   & 0.364 & 0.250 & 0.806  \\ \hline
\end{tabular}
\label{tab10}
\end{center}
\end{table}

From the data in Table \ref{tab10}, it can be seen that the GKT model is effective on the dataset \textit{Subject-Math}, with MSELoss up to 0.041, RMSE score up to 0.202, MAE score up to 0.041, and ACC score up to 0.960.

And the results prove that the GKT model has a huge advantage when predicting on datasets with clear and logical knowledge point relationships.


\section{Conclusions} \label{CON}
Based on three classic knowledge tracing models, namely DKT, DKVMN, and GKT, this paper leverages the publicly available \textit{ASSISTment-2009-2010-skill} dataset and private datasets \textit{Object}, \textit{Subject} from Changshui High School's monthly exams for sophomore class students, which include answer data for both objective and subjective test questions across eight academic disciplines, to implement a unified knowledge tracing model for both objective and subjective test questions. Initially, in the context of objective test questions, a binary classification approach is adopted for model training. Experiments conducted on three benchmark models using real data demonstrate that the models exhibit good knowledge tracing and prediction capabilities for objective test questions. Given the characteristics of subjective test questions, the classification task is transformed into a regression task. By uniformly modifying the model training approach, adding branch networks, and optimizing the method of question encoding, knowledge tracing for subjective test questions is achieved. When the uniformly adjusted benchmark models are applied to subjective test questions, the prediction results are remarkable. The unified knowledge tracing model for both objective and subjective test questions studied in this paper can fully utilize exam data, combine objective and subjective test questions, and provide a qualitative and quantitative comprehensive assessment of learners' knowledge states, thereby offering reliable support for personalized teaching. Finally, this paper designs experiments, presents and analyzes the experimental results. The main limitation of this study lies in the adjustment method of the model and the prediction method for subjective test questions in subjective test scenarios, which still need to be improved. The current adjustment method of the model is mainly based on the encoding method of the original model, which simply encodes the knowledge points and answer scores of subjective test questions, and then changes the loss function to MSELoss commonly used in regression tasks. However, the prediction performance is lower than that of objective test questions.

\bibliographystyle{IEEEtran}
\bibliography{Mybib,IEIR}

\end{document}